\documentstyle[aps,prb,epsf,floats]{revtex}

\newcommand{\be}{\begin{equation}}
\newcommand{\ee}{\end{equation}}
\newcommand{\bi}{\bibitem}
\newcommand{\D}{\Delta}

\newcommand{\rDT}{r_{\scriptscriptstyle \Delta T}}
\newcommand{\rDJ}{r_{\scriptscriptstyle \Delta J}}
\begin{document}
%\twocolumn[\hsize\textwidth\columnwidth\hsize\csname@twocolumnfalse%
%\endcsname
\title{Chaos and Universality in a Four-Dimensional Spin Glass}
\author{Muriel Ney-Nifle\cite{perm:add}}
\address{Department of Physics, University of California\\
Santa Cruz CA 95064}
\maketitle
\begin{abstract}
We present a finite size scaling analysis of Monte Carlo simulation
results on a four dimensional Ising spin glass. We study chaos with
both coupling and temperature perturbations, and find the same
chaos exponent in each case. Chaos is investigated both at the
critical temperature and below where it seems to be more efficient
(larger exponent). Dimension four seems to be
above the critical dimension where chaos with temperature is no more
present in the critical region. Our results are consistent with the Gaussian
and bimodal coupling distributions being in the same universality class.
\end{abstract}

\section { Introduction}
While the nature of the spin glass phase is still controversial, there is
a property, namely static chaos, that emerges as a common feature of
different spin glass models.
This chaotic behavior 
has been studied within mean-field theory \cite{p,k,fn}, 
scaling \cite{bm} or droplet \cite{fh}
theories and using a real space renormalisation-group approach \cite{bb,nh}. 
Chaos means that the frozen random equilibrium state of the spin
glass phase is completely reorganised by a small change in an external
parameter, such as temperature (T) or magnetic field. Chaos with respect
to a slight change in the the couplings, a so-called random perturbation,
has also been studied. 
In the case of a temperature preturbation, for example, the
spin-spin correlation varies chaotically with T, the larger
the distance between the spins, the bigger this effect.
The temperature scale at which the correlation function varies is
\be
\D T_{L} \sim L^{-\zeta},
\label{zeta0} 
\ee
which defines the Lyapunov or chaos exponent $\zeta$. 
The existence of such a scaling
and of $\zeta$ has been shown within the models quoted
above and a summary of the findings will be presented later in this
introduction. Here we study chaos by Monte Carlo (MC) simulations of a
four-dimensional spin glass, and we consider the Edwards-Anderson model
with Gaussian coupling distribution.
Some of the questions adressed in this paper are: 

$\bullet$ Are all perturbations equivalent ?
Chaos with two different random perturbations and a temperature
change have been studied at the
critical temperature and are found to give a unique exponent, within
the uncertainties. The amplitudes of the effect
are not the same, however, and a temperature perturbation is more difficult
to see numerically than the random perturbations.

$\bullet$ Are there different chaos exponents at the critical temperature and
in the spin glass phase ? 
We get rather different
values of $\zeta$, at $T_c$ and below. This seems to support previous findings
that chaos is more effective in the spin glass phase (larger exponent)
\cite{nh}. 

A brief reminder is needed to compare these results with
ones of other models. 
Chaos in spin glass phase was first pointed out
in mean-field theory by Parisi \cite{p}, and later confirmed \cite{fn}.
A loop-expansion around Parisi's solution for
dimensions $d\geq 8$ allowed Kondor et al. \cite{k} to show chaos with magnetic
field and also with temperature. Two distinct exponents $\zeta$ resulted
while the temperature perturbation has a smaller effect.

For low dimensional systems, chaos has been extensively studied within
the scaling theory of Bray and Moore \cite{bm}
and the droplet theory of Fisher and Huse \cite{fh} which are based on 
a real space renormalisation group approach. The latter permits one to determine
chaos exponents for arbitrary perturbations and in various points of
the phase diagram. Chaos has been shown to be characteristic of each
fixed point of the diagram \cite{nh}. In particular, one gets different exponents
at the critical temperature ($T_c$)
(between spin glass and paramagnetic phases)
and in the ordered phase ($T<T_c$), in dimension three.
One can argue that there is a critical dimension above which there is {\em no} chaos 
with temperature anymore at $T_c$ \cite{nh}.
In the Migdal-Kadanoff (MK) framework which is the renormalisation scheme that
has been mainly used, dimension four is above this critical dimension. 
This leads to an additional question for our study:

$\bullet$ What is this critical dimension above which chaos with
temperature no longer exists at the critical temperature?
Our Monte Carlo simulation indicates that four is close to but higher than
this critical dimension.

Moreover, within MK, the exponents from a temperature perturbation
or a random perturbation have to be the same, keeping everything else fixed,
(in agreement with this paper's results).
Finally, a magnetic field perturbation has been less studied but seems to lead
to a different exponent\cite{n}. As it was the case with mean-field approach,
chaos with temperature is less effective (smaller exponent)
than with magnetic field. A major difference, however, is that a magnetic
field destroys the spin glass phase and chaos is studied in the
paramagnetic phase close to zero field while, in mean field, one stays
below a non-zero critical field. 

MC simulations
of the Edwards-Anderson model allow comparisons with
both previous approaches \cite{mpr}. Ritort's simulations \cite{r}
of the mean-field version of this
model confirms the analytical results of Kondor. In dimension two, where
$T_c$ is zero, we have been able to observe chaos with
both temperature and random perturbation and found a unique exponent
\cite{ny}.
We are now interested in
higher dimensions where there is a spin glass phase. Since
dimension four has a clear transition in contrast to dimension three \cite{ky},
and also because there has already been some studies of chaos
for $\pm J$ random couplings \cite{afr}, we focus on a Gaussian coupling four-dimensional
system. Thus we are able to answer to the question :

$\bullet$ Are exponents the same for different coupling
distributions ?
The results are consistent with the Gaussian (this paper)
and bimodal (Ritort et al. \cite{afr}) distributions lying in the same universality class.

Chaos with temperature or magnetic field are believed to explain cycling temperature \cite{rv}
or magnetic field \cite{dg} experiments on spin glasses. Such an experiment has been done recently 
on a disordered ferromagnet\cite{jm}. This can be understood whithin renormalisation group 
approach\cite{nh} where chaos is shown to be present when the coupling distribution
is shifted towards ferromagnetic couplings. It could be interesting to check that
chaos can also be seen with Monte Carlo simulations. 

\section { The Model}

The Hamiltonian (for Ising spins $\lbrace S_i \rbrace$ 
and nearest neighbor couplings $\lbrace J_{ij} \rbrace$), is
\be
{\cal H} = -\sum_{\langle i, j\rangle} J_{ij} S_i S_j \ ,
\ee
where the couplings are drawn from a Gaussian distribution with
zero mean and variance $\overline{ J_{ij}^2 }$ equal to unity. 
The spins lie on a four-dimensional cubic lattice of linear size $L$ 
with periodic boundary conditions.
As in previous studies \cite{ky,afr,by,ny}, the basic quantity is the
replica overlap between two copies (replicas $a$ and $b$) of the system 
\be
q_{ab} = {1 \over L^2} \sum_{i=1}^{L^2} S_i^{(a)} S_i^{(b)} \ .
\label{qab}
\ee 
From this the Binder ratio is computed,
\be
g   \equiv  {1 \over 2} 
\left[ 3 - 
{ \langle q^4 \rangle \over \langle q^2 \rangle^2 } 
\right]
\label{g} 
\ee
where $\langle \cdots \rangle$ denotes both the average over disorder and the
statistical mechanics (Monte Carlo) average.
It is dimensionless and has a finite size scaling which allows for
calculation of the critical temperature
\be
g  =   \tilde{g} \left(L^{1/\nu} (T-T_c) \right) .
\label{gscale}
\ee

A similar approach has been introduced by Ritort \cite{afr} to investigate chaos
with random perturbation. One has now to compare two copies with correlated
coupling sets. We obtain the 
perturbed couplings $\lbrace
J^\prime_{ij}\rbrace$ from the unperturbed ones 
$\lbrace J_{ij} \rbrace$ in two
ways. First,
\be
\label{dJ}
(1) \ \ J^\prime_{ij} =
{{J_{ij} + x_{ij} \D J}\over {\sqrt{1+{\Delta J}^2}}} \ ,
\label{DJ1}
\ee
where $x_{ij}$ is a Gaussian random variable with 
zero mean and unit variance. Secondly, we consider the perturbation
that changes the sign of a small fraction of the couplings,
\be
(2) \ \ J^\prime_{ij} = - J_{ij} \ \ {\rm with \ probability \ p}
\label{DJ2}
\ee
and $J^\prime_{ij} = J_{ij}$ otherwise. With $\pm J$ coupplings, one can
only study case $(2)$ \cite{afr}. 
 In order to compare both perturbations, one can determine how the two
sets of couplings are correlated. One gets
\begin{eqnarray}
\nonumber
(1) \ \ \ \overline{J_{ij} J^\prime_{ij}} & \simeq & 1 - {{\D J^2} \over 2} \\
(2) \ \ \ \overline{J_{ij} J^\prime_{ij}} & = & 1 - 2 p \ .
\label{correlation}
\end{eqnarray}
where $(1)$ is expanded for small $\D J$.

A measure of the spin reorganisation under small perturbation
is the chaos parameter \cite{afr,ny}
\be
\rDJ \equiv 
{\langle q_{\scriptscriptstyle JJ^\prime}^2 \rangle \over \langle
q_{\scriptscriptstyle JJ}^2 \rangle} ,
\label{rJ}
\ee
where $q{\scriptscriptstyle JJ^\prime}$ is given by Eq.(\ref{qab})
with two copies $a$ and $b$
having now slightly different couplings. Note that 
$q{\scriptscriptstyle JJ} = q{\scriptscriptstyle JJ^\prime}$
since $\lbrace J^\prime_{ij}\rbrace$,
and the $\lbrace J_{ij}\rbrace$ have the same distribution.
The scaling of this quantity, at fixed temperature, leads to a chaos exponent
for each random perturbation $(1)$ and $(2)$,
\be
\rDJ = \tilde{r}_{\scriptscriptstyle \D J}\left( L^\zeta \D J \right)
\label{rJscale} 
\ee

For a temperature perturbation, one can follow the same pathway 
\cite{ny} and
define a parameter $\rDT$ like in Eq. (\ref{rJ}) 
%except that  now
%$T_+$ and $T_-$ have to be subsituted to $J$ and $J'$
\be
\rDT \equiv 
{ \langle q^2_{\scriptscriptstyle T_+ T_-} \rangle \over
\sqrt{ \langle q^2_{\scriptscriptstyle T_+ T_+} \rangle \langle
q^2_{\scriptscriptstyle T_- T_-} \rangle} } ,
\label{rT}
\ee
where $q_{\scriptscriptstyle T_+ T_-}$
is given by (\ref{qab}) with two copies $a$ and $b$
having slightly different temperatures, $T_-$ and $T_+$,
which are equally shifted from a reference
temperature $T$, that is, 
\be
T_{\pm} = T \pm \D T / 2 \ .
\label{DT}
\ee
Finally, an analoguous finite size ansatz is applied to this parameter
\be
\rDT = \tilde{r}_{\scriptscriptstyle \D T}\left( L^\zeta \D T \right) ,
\label{rTscale} 
\ee
which could lead to another chaos exponent $\zeta$ (although we use the same symbol). 

\section { Results and Discussion}

\subsection { Chaos at $T_c$}

A plot of $g$ as a function of $T$ for various sizes $L$
shows an intersection at $T_c = 1.8$, see Fig.1.
Knowing this
permits one to determine the critical exponent $\nu$ from a plot
of $g$ as a function of $(T-T_c) L^{1/\nu}$, as usual \cite{by,ky}.
In such a plot,
see Fig.2 , a collapse of all data
to a single curve is obtained for 
\be
T_c = 1.8 \pm 0.05\ ,\ \ \nu = 0.87 \pm 0.15 \ .
\label{nu}
\ee
This is in agreement with previous MC simulations\cite{mpr,by}.

Scaling plots of the chaos parameter for the three
perturbations are shown in Fig. 3, 4 and 5
where the reference temperature is precisely $T_c$.
In the case with random perturbations, Eqs. (\ref{DJ1}, \ref{DJ2}),
and temperature change given by Eq. (\ref{DT}) with $T=T_c$, one gets 
\begin{eqnarray}
\nonumber
%{\rm chaos \ with} \ \D J, {\rm case} \ (1) \ : & \zeta \ = \ 0.85 \pm 0.1 \\
%\nonumber
{\rm chaos \ with} \ \D J \ : & \zeta \ = \ 0.85 \pm 0.1 \\
{\rm chaos \ with} \ \D T \ : & \zeta \ = \ 0.95 \pm 0.2 \ .
\label{zeta}
\end{eqnarray}
One can check that exponents from both random perturbations are equal
(see Fig. 3 and 4). 
The data for chaos with $T$ in Fig.5 has larger statitical errors and also
deviates less from unity; in other words, the amplitude of chaos with $T$
is smaller which is also the case with other simulations \cite{afr} and mean-field
calculations \cite{k}. 

We conclude that the exponents for chaos with $\D J$ and $\D T$ at $T_c$, given by
Eq.~(\ref{zeta}), are equal within the error bars.
%whatever the type of random perturbation used. 
We use this to compare results from several sources
for chaos with $\D J$ and $\D T$ together in Table I.

\begin{table}
\begin{center}
\caption{ 
Values of the chaos exponent is shown for various dimensions. In some cases
we have $\zeta$ both in the spin glass phase, $T<T_c$, and in the critical region, at $T_c$
(numbers in brakets). The models
are Migdal Kadanoff Renormalisation group calculations (MKRG), mean-field
expansions around Parisi's solution, Monte Carlo simulations for Gaussian (G)
coupling distributions (this paper's
results in dimension four, and our previous results in dimension two 
\protect\cite{ny}), and bimodal ($\pm J$) coupling distribution 
(Note that Ritort et al.
used another symbol $\lambda \equiv 2/\zeta$)
and finally exact ground state numerical calculations. The chaos exponent has been
determined both with random and temperature perturbations except when
it is mentioned $\D J$ or $\D T$.
}
\label{table1}
\begin{tabular}{|c|c|c|c|c|}
dimension $\rightarrow$  & 2 & 3 & 4 & $\geq$8\\
model $\downarrow$ & $T_c=0$ & $T<T_c (T_c)$ & $T<T_c (T_c)$ & $T<T_c$ \\ \hline\hline
MKRG\cite{nh} & 0.73 & 0.73 (0.57) & 0.73 (0.53) & - \\
%---- $1/\nu$ & ? & - (0.35) & - (0.68) & - \\
Mean-field\cite{k,fn} ($\D T$) & - & - & - & 1.0 \\
Monte Carlo\cite{afr} ($\pm J$) & - & - & 1.0 (0.75) & - \\
Monte Carlo (G) & 1.0 & - & 1.2 (0.85) & - \\
Ground State\cite{rsb} ($\D J$) & 0.95 & - & - & - \\
\end{tabular}
\end{center}
\end{table}

In order to define chaos with $\D T$ in the critical
region it is necessary that $\zeta > 1/\nu$ so that the typical temperature
interval on which the spin correlation varies, $\D T_{L}\sim L^{-\zeta}$,
is smaller than the critical temperature range given by $|T-T_c|\sim L^{-1/\nu}$.
Both exponents depends on dimension. In our four dimensional system
$1/\nu = 1.15$, and
the inequality does not seem to be satisfied (or there might be
an equality) ruling out chaos with $T$ in the critical region. 
This is also the case with MKRG which gives $1/\nu = 0.68$ (to be compared
with $\zeta=0.53$ in Table I). 
According to MKRG, chaos at the critical temperature
{\em is} present in dimension three, $1/\nu = 0.35$; this is not
yet confirmed by Monte Carlo simulations.

Finally, one can see from 
Table I, in dimension four (numbers in brakets), that 
the exponents for Gaussian and $\pm J$ distributions are equal
(error bars are of order $\pm 0.1$).

\subsection { Chaos in the spin glass phase}

We also calculate the chaos exponent in the spin glass phase. From a previous
numerical result of Young et al. \cite{rby}, temperature $T=1.4$ and sizes $L \geq 3$
seems to probe the ordered low temperature phase. The data for $T=1.4$ is shown
in Fig. 6 from which we estimate
\be
{\rm chaos \ with} \ \D J, {\rm case} \ (1) \ : \ \ \zeta \ = \ 1.2 \pm 0.1 \ .
\ee
This exponent is larger (although we cannot rule out an equality)
than the one at $T_c$. This
indicates that chaos is more efficient at low temperature in
agreement with other sources quoted in Table I, see
the dimension four column.

\subsection { Conclusions}

We find that the Gaussian and bimodal distributions seems to be in the same
universality class (in contrast to numerical calculations of 
other critical exponents \cite{c}). 

Our results on chaos in four dimensions (this paper) 
and in two dimensions \cite{ny}
are in qualitative agreement with those of the 
real-space-renormalisation-group approach\cite{nh}. Some of the
results obtained in this approach\cite{nh} and observed numerically are the following. First, 
chaos is also present in the paramagnetic phase (e.g., in dimension two\cite{ny}). Second, a
temperature perturbation generates a random perturbation and thus 
gives the same chaos exponent (confirmed numerically in both two 
and four dimensions).
Moreover, chaos
with temperature occurs inside the critical region if the dimension is
smaller than a limit dimension \cite{nh}, called $d_+$, with $3 < d_+ < 4$
(in agreement with this paper's results).
The chaos exponent is smaller at $T_c$ 
than in the spin glass phase (see Table I). 
More generally, it has been argued \cite{nh} that 
a chaos exponent can be assigned
to each fixed point of each random system, and can appear in physical
quantities when a perturbation of the thermodynamic parameters couples to
a random perturbation. The disordered ferromagnet is one example \cite{jm}, and
it could be interesting to perform a Monte Carlo simulation of a spin glass
with random ferromagnetic couplings.

\acknowledgments
This work was supported by a NATO fellowship and 
by the Centre National de la Recherche Scientifique, and
was performed during my stay at the Physics Dept
of UC, Santa Cruz where I benefited from
disussions with Peter Young.

\begin{figure}
\epsfxsize=\columnwidth\epsfbox{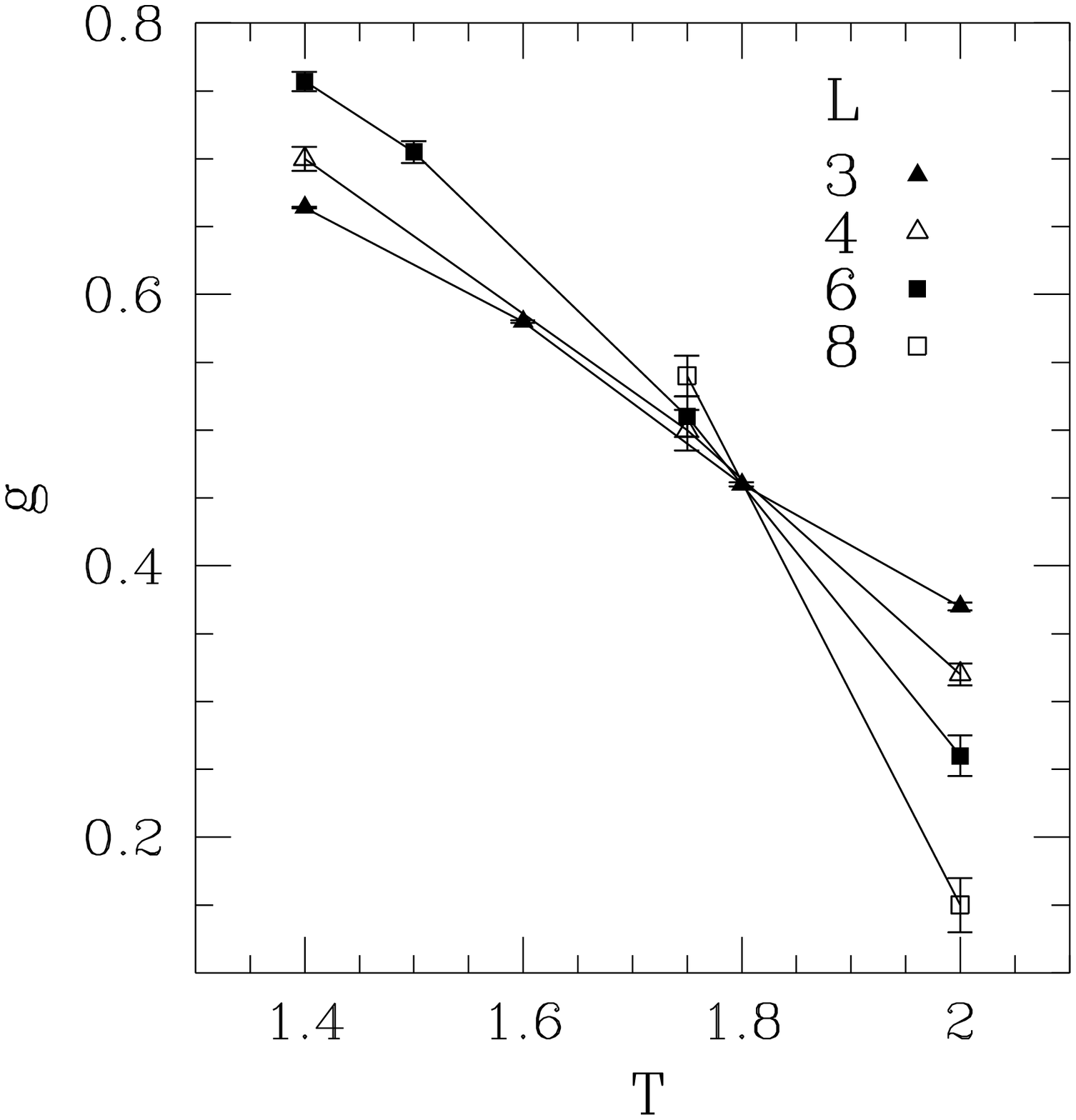}
\caption{
Data for g against T. From the intersection, $T_c$ is
estimated to be $1.8$.
}
\label{plot:gT}
\end{figure}
 
\begin{figure}
\epsfxsize=\columnwidth\epsfbox{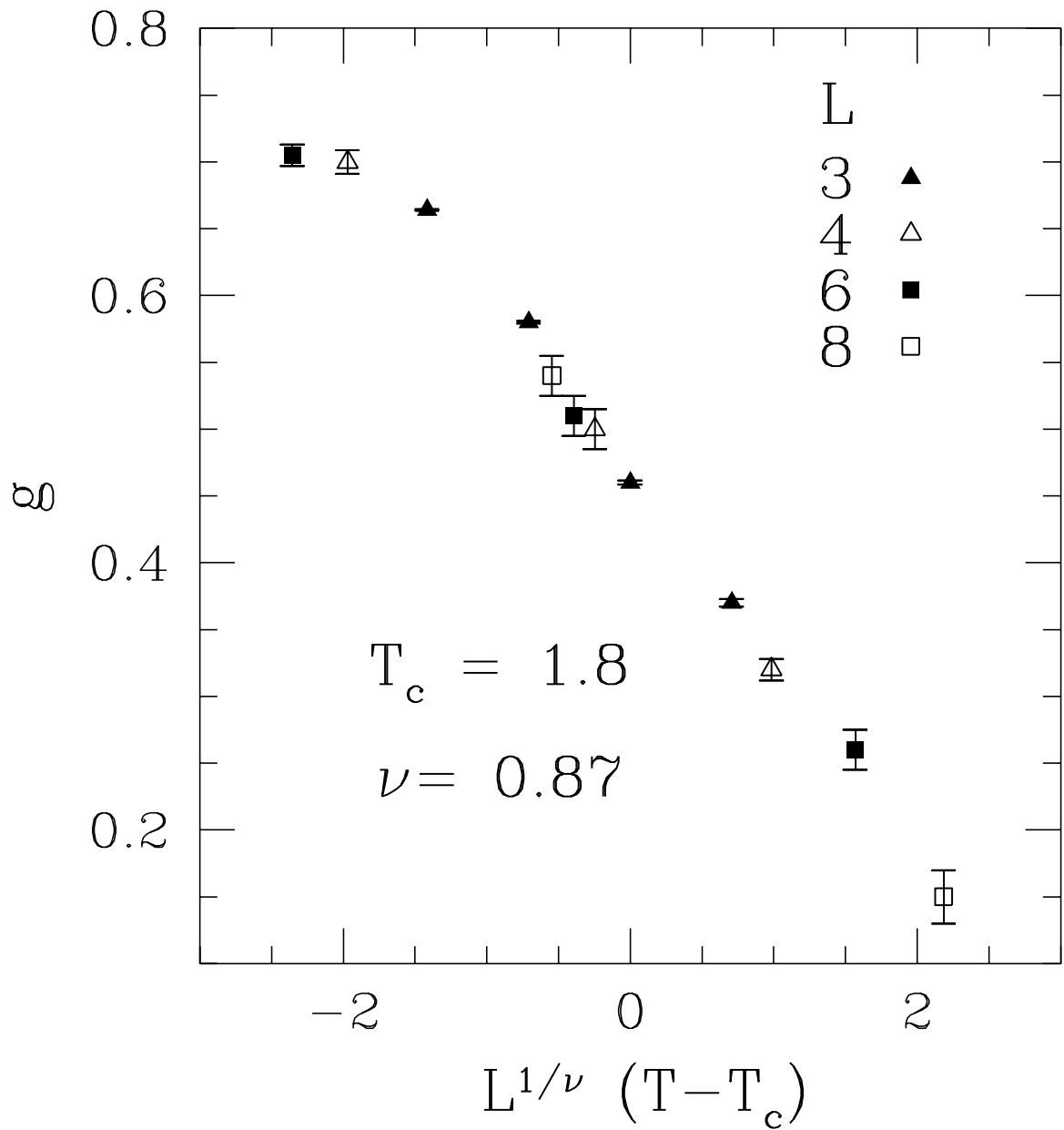}
\caption{A scaling plot of the data in Fig.1 scaled according to
Eq. (\protect\ref{gscale}) which gives $\nu=0.87 \ (1/\nu=1.15)$.}
\label{plot:gnu}
\end{figure}

\begin{figure}
\epsfxsize=\columnwidth\epsfbox{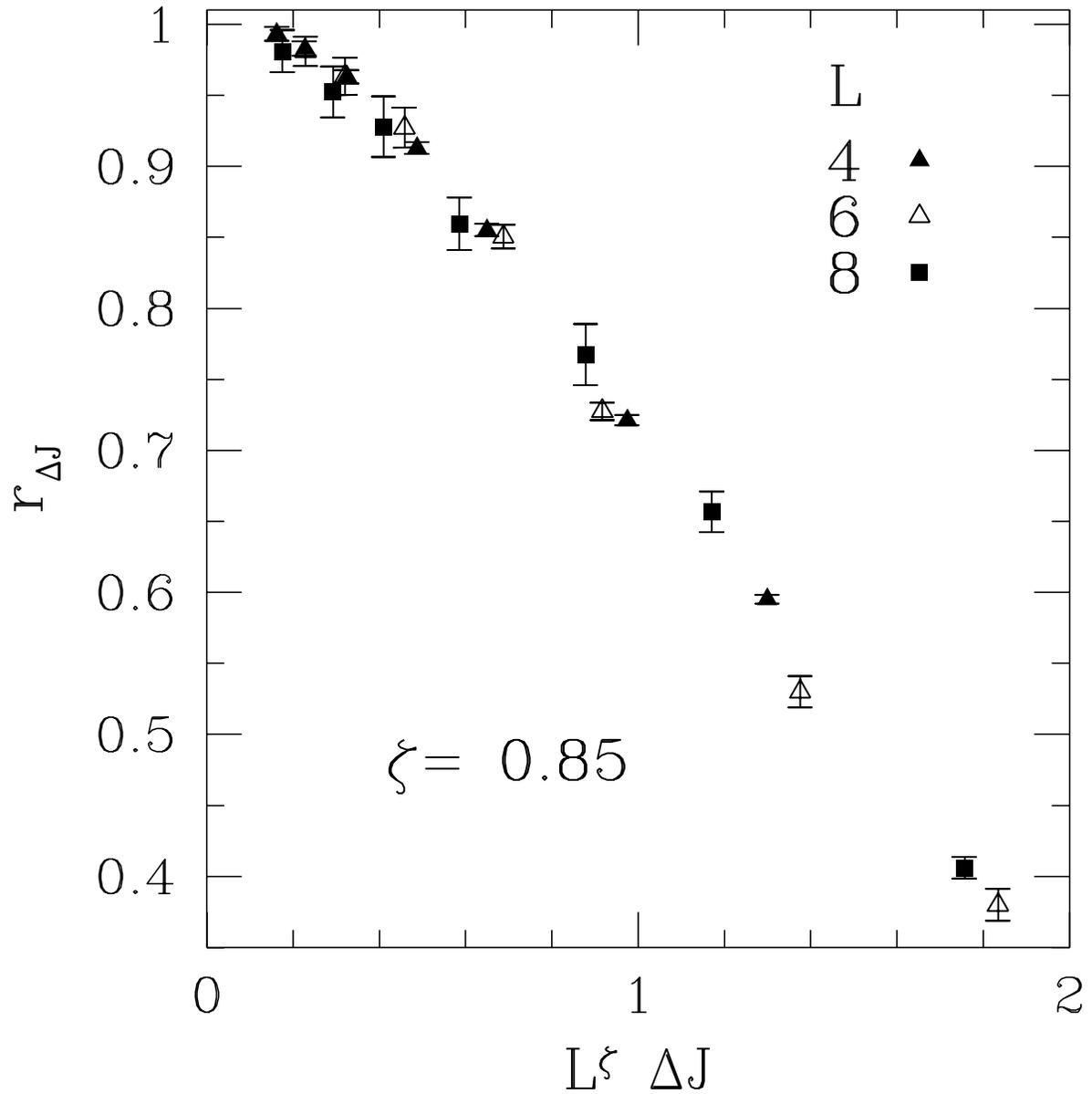}
\caption{
A scaling plot of $r_{\scriptscriptstyle \Delta J}$ with random perturbation, 
case $(1)$, {\em at} $T_c$.
The perturbation lies in the range $0.05 - 0.4$. Trying
different values, our best estimate is $\zeta=0.85$.
}
\label{plot:rJ1}
\end{figure}

\begin{figure}
\epsfxsize=\columnwidth\epsfbox{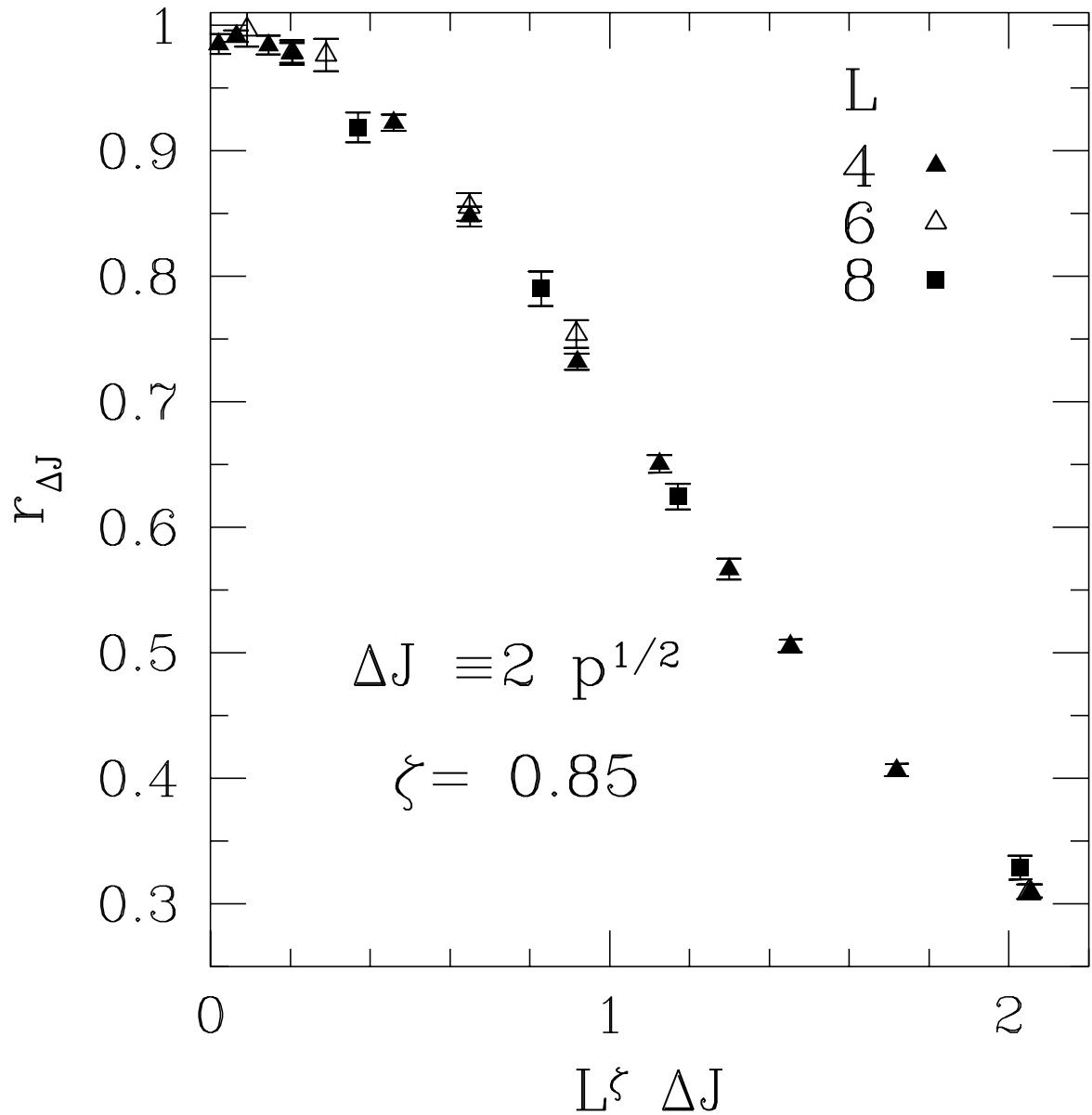}
\caption{
A scaling plot of $r_{\scriptscriptstyle \Delta J}$ with random perturbation, case $(2)$,
{\em at} $T_c$.
To compare these data with Fig. 3, we use Eq. (\protect\ref{correlation})
and define $\Delta J \equiv 2 p^{1/2}$. 
The perturbation $p$ lies in the range $0.001 - 0.1$
which is $\Delta J = 0.063 - 0.63$.
The chaos exponent is $\zeta=0.85$.
}
\label{plot:rJ2}
\end{figure}

\begin{figure}
\epsfxsize=\columnwidth\epsfbox{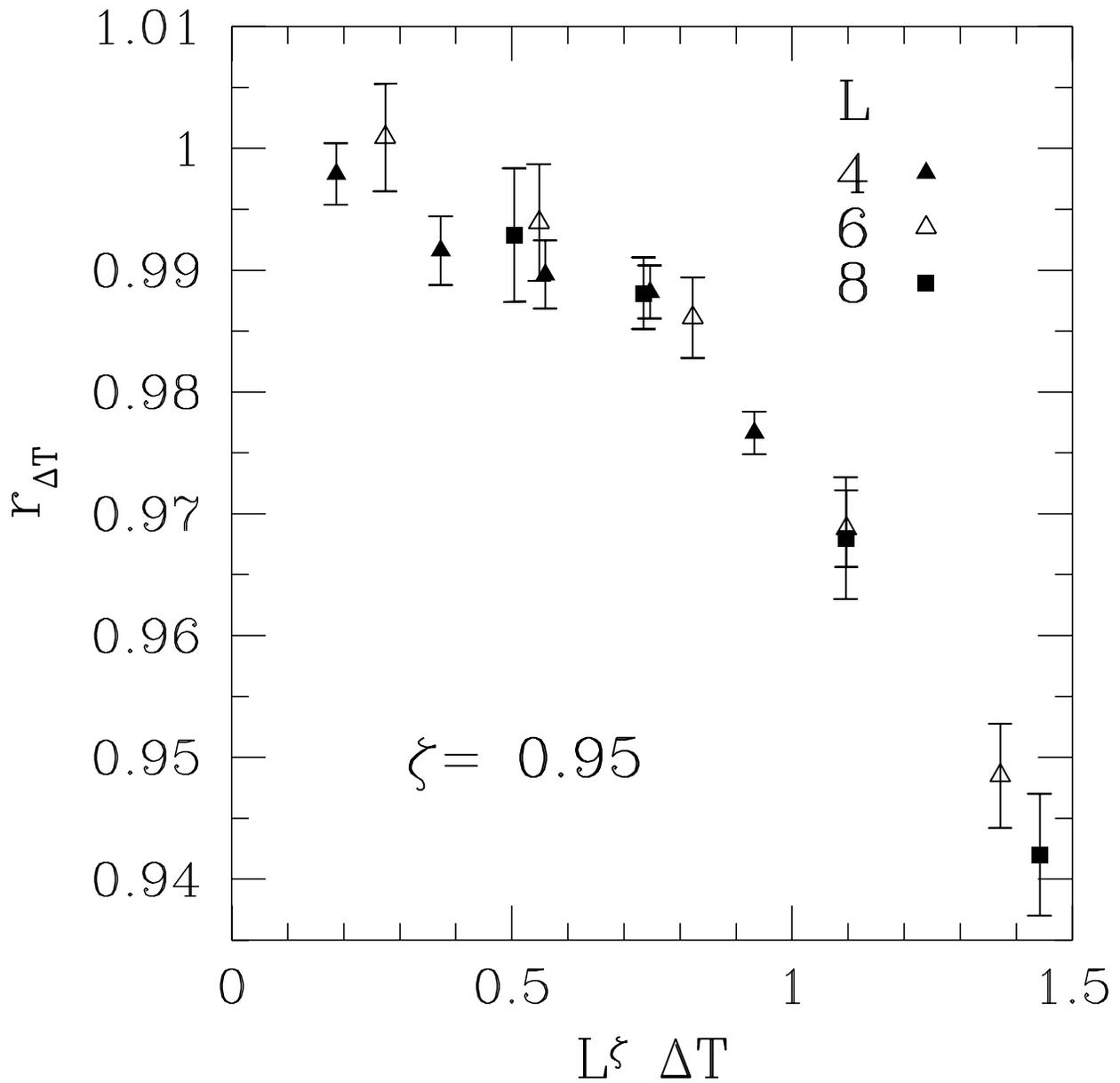}
\caption{
A scaling plot of $r_{\scriptscriptstyle \Delta T}$ {\em at} $T_c$.
The perturbation, $\Delta T$, lies in the range $0.05 - 0.25$.
The chaos exponent is $\zeta=1.0$.
}
\label{plot:rT}
\end{figure}

\begin{figure}
\epsfxsize=\columnwidth\epsfbox{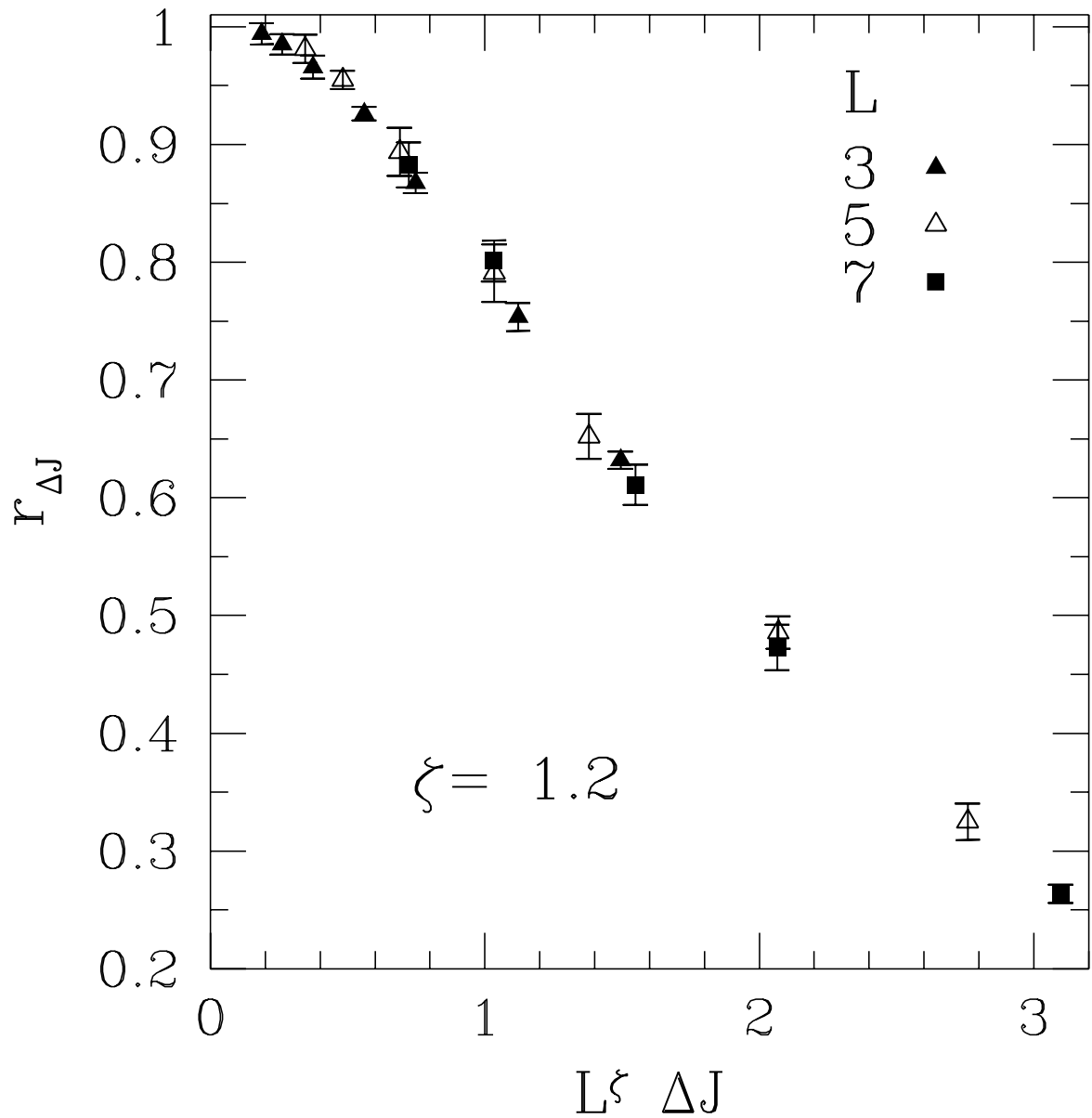}
\caption{
A scaling plot of $r_{\scriptscriptstyle \Delta J}$ with random perturbation, 
case $(1)$, at $T=1.4$ in the spin glass phase.
The perturbation lies in the range $0.05 - 0.4$.
The chaos exponent is $\zeta=1.2$.
}
\label{plot:rJ}
\end{figure}

\end{document}